\documentclass[oldversion]{aa} % double column 
\usepackage{graphicx}
\usepackage{aalongtable}
\usepackage{txfonts}
\usepackage{rotating}
\usepackage{lscape}
\newcommand{\gsim}{\;\lower.6ex\hbox{$\sim$}\kern-7.75pt\raise.65ex\hbox{$>$}\;}
\newcommand{\lsim}{\;\lower.6ex\hbox{$\sim$}\kern-7.75pt\raise.65ex\hbox{$<$}\;}

\begin{document}
\title{Homogeneous abundance ratios of hydrostatic and explosive alpha-elements
in globular clusters from high resolution optical spectroscopy\thanks{Based on observations  collected at ESO telescopes under
programmes (see Appendix)}
%\thanks{Full Table 3 and Table A.1 in Appendix A are only available at the CDS
%via anonymous ftp to cdsarc.u-strasbg.fr (130.79.128.5) or via http://cdsarc.
%u-strasbg.fr/viz-bin/cat/J/A+A/??/??}
 }

\author{
Eugenio Carretta\inst{1}
}

\authorrunning{E. Carretta}
\titlerunning{Abundances of hydrostatic and explosive alpha-elements}

\offprints{E. Carretta, eugenio.carretta@inaf.it}

\institute{
INAF-Osservatorio di Astrofisica e Scienza dello Spazio di Bologna, Via P. Gobetti
 93/3, I-40129 Bologna, Italy}

\date{}

\abstract{Galactic globular clusters (GCs) were born shortly after the Big Bang.
For such old stellar systems the initial mass function (IMF) at the high mass
regime can never be observed directly, because stars more massive than about 1
M$_\odot$ have evolved since longtime. However, the hydrostatic to explosive
$\alpha-$element ratio (HEx ratio) offers a way to bypass the lack of observable
high mass stars through the yields that massive stars released when exploding as
supernovae, incorporated in the stars we presently observe in GCs. The HEx
ratio measures the percentage of high mass stars over the total number of stars
exploding as supernovae and it is an efficient probe of the ephemeral first
phases of the GC evolution. We exploited a recently completed survey 
to assemble a dataset of very homogeneous abundances of $\alpha-$elements in 27
GCs from [Fe/H]$\sim -2.4$ to $\sim -0.3$ dex. In agreement with previous
results from APOGEE, we confirm that the HEx ratio is indistinguishable for GCs
formed in situ and accreted in the Galaxy, and that this ratio decreases with
increasing metallicity. However, we posit that this trend is better explained by
a metallicity-dependent IMF  deficient in the highest mass stars at high
metallicity, as corroborated by the declining [O/Mg] ratio as a function of the
[Mg/H] ratio. At odds with the previous analysis based on APOGEE data, we detect
an anti-correlation of HEx ratio with both present day and initial GC masses. Finally,
we hypothesise that in that analysis, the stars of the GC M~54 were probably
confused with stars in the core of the Sagittarius dwarf galaxy, where the
cluster is presently immersed.}

\maketitle

\section{Introduction}

Oxygen follows hydrogen and helium as the most abundant elements in the
Universe. Reactions with successive captures of He nuclei ($\alpha$ particles)
on Ne synthetise Mg, Si, Ca, and Ti (e.g. Burbidge et al. 1957), collectively
known as $\alpha-$elements. These elements are enhanced in metal-poor stars
(see Wallerstein 1962) and in particular in globular clusters (GCs) of the Milky
Way (MW). This occurrence is explained with the interplay between star formation and
the lifetime of stars in different mass ranges (Tinsley 1979). Massive stars
evolve quickly, leading to a quick (prompt) release of copious amounts of
$\alpha-$elements in core-collapse supernovae (CCSNe), well before the bulk of 
contribution by type Ia SNe, whose precursors evolve on a longer (delayed)
timescale, may lower the [$\alpha$/Fe]\footnote{We adopt the usual spectroscopic
notation, $i.e.$  [X]= log(X)$_{\rm star} -$ log(X)$_\odot$ for any abundance
quantity X, and  log $\epsilon$(X) = log (N$_{\rm X}$/N$_{\rm H}$) + 12.0 for
absolute number density abundances.} ratio raising the iron content of the
yields. The level of $\alpha-$elements measured in a stellar system is then a
probe of the efficiency of the system to produce and retain yields from massive
stars before the onset of the bulk of SNe Ia.

However, the dominant nucleosynthetic channels are different for different
species belonging to this group. The synthesis of O and Mg is dominated by
hydrostatic burning in the outermost shells of the most massive stars (in the
range 15-30 M$_\odot$) before being released by CCSNe (e.g. Weinberg et al. 2019
and references therein). As a consequence, they
are dependent on the mass of the star, with yields increasing with increasing
progenitor mass, but only a weak dependence on metallicity at fixed stellar mass (e.g. Woosley
and Weaver 1995, Andrews et al. 2017, Griffith et al. 2019). 
Explosive $\alpha-$elements like Si, Ca, and Ti are produced in shells lying
closer to the core of massive stars. This makes their yields relatively
independent of stellar mass when they are released by SNe, predominantly by
CCSNe, with a lower, but significant contribution from SNe Ia (e.g. Nomoto et
al. 1984).

These different properties means that the comparison of $\alpha-$elements
produced in hydrostatic burning (O, Mg) to the abundance of explosive
$\alpha-$elements (Si, Ca, Ti) is an efficient way to sample the high mass end
of the initial mass function (IMF) of a given stellar population (e.g. McWilliam et
al. 2013, Carlin et al. 2018) through the analysis of the yields incorporated in
the stars we are presently observing. For GCs this is an important probe, since
stars more massive than $\sim 1$ M$_\odot$ are since longtime evolved off the
main sequence and are difficult to observe.

The first and most straighforward applications of this method were both for
the Sagittarius dwarf galaxy (Sgr dSph). Using the [Mg/Ca] ratio as example of
the hydrostatic-to-explosive abundance ratio of $\alpha-$elements McWilliam et
al. (2013) concluded that the decline of this ratio with increasing metallicity
suggests the evidence of an IMF deficient in the highest mass stars. Carlin et
al. (2018), who coined the term HEx ratio, reached similar conclusion by
studying this ratio in 42 stars in the Sgr tidal stream. Both studies were based
on optical high resolution spectroscopy. 

The same approach is made more complicated in GCs by the presence of multiple
stellar populations (MPs). Whereas most heavy elements show a high degree of
homogeneity in GCs (e.g. Gratton et al. 2004), MPs are distinct by variations
of the abundances of light elements, modified by proton-capture
reactions with respect to the primordial level established in GCs by SNe (see
Gratton et al. 2012, 2019 for comprehensive reviews). In the network of
proton-capture reactions in H-burning at high temperature (e.g. Langer et al. 
1993) the hydrostatic $\alpha-$elements O and Mg are both consummed to produce
N and Al. Oxygen may be depleted even down to [O/Fe]$\sim -1$ dex starting from
the mean enhancement (0.3-0.4 dex) typical of GCs, and its variations are observed
in almost every GCs in the MW (e.g. Carretta et al. 2010a). Depletion of Mg
abundances is much more moderate, and significant variations are only observed
in GCs that are metal-poor, massive, or both (Carretta et al. 2009a, M\'esz\'aros
et al. 2020). Also the explosive elements Si and Ca are affected by the MPs
phenomenology. In the case of Si, the changes in abundances are usually small,
since the main mechanism is actually a leakage on Si from the Mg-Al cycle (e.g.
Arnould et al. 1999, Yong et al. 2005). Variations in Ca, detected as small yet
observable excesses with respect to unpolluted field stars of similar
metallicity, are much more rare, being found in about 10\% of all the surveyed
GCs (Carretta and Bragaglia 2021).

A large survey of HEx ratios in GCs, as well as in halo substructures and
satellites galaxies of the MW,  was recently done by Horta and Ness (2025;
hereinafter HN25) exploiting the APOGEE survey DR17. They used the
unpolluted stellar population in GCs selecting N-poor/C-rich and 
Al-poor/Mg-rich stars
and they favour an alternative scenario explaining the decreasing trend of HEx
ratios in GCs with the metallicity with a delayed contribution of SNe Ia,
leading to a higher contribution of both iron and explosive $\alpha-$elements.

In the present work we are taking advantage of a recently
completed survey of Mg, Si, Ca, and Ti from high resolution optical spectroscopy
in 16 GCs (Carretta 2026). We provide a description of the available data in
Section 2, compute the HEx ratios for GCs in Section 3, and compare our results
with those from the APOGEE survey (Section 4). Finally, we discuss and summarise
our findings in Section 5.

\section{Datasets}

\begin{table*}
\centering
\caption{Average abundances of hydrostatic and explosive $\alpha-$elements}
\scriptsize
\begin{tabular}{lrrrrrr}
\hline

GC     &    [O/Fe]       &   [Mg/Fe]       &   [Si/Fe]       & [Ca/Fe]         & [Ti/Fe]~{\sc i}& [Fe/H]     \\
       & n~~~ mean~ rms  & n~~~ mean~ rms  & n~~~ mean~ rms  & n~~~ mean~ rms  & n~~~ mean~ rms   & mean~ rms  \\
\hline
0104   &115 +0.152 0.179 &147 +0.530 0.076 &147 +0.433 0.061 &147 +0.315 0.024 &146 +0.394 0.036 &-0.768 0.054  \\
P only & 31 +0.289 0.083 & 31 +0.533 0.081 & 31 +0.399 0.058 & 31 +0.310 0.023 & 31 +0.394 0.033 &              \\
\hline     
0288   & 70 +0.132 0.257 &107 +0.469 0.045 &112 +0.389 0.029 &111 +0.402 0.045 &106 +0.283 0.040 &-1.305 0.054  \\
P only & 23 +0.286 0.196 & 22 +0.465 0.043 & 23 +0.379 0.035 & 23 +0.394 0.035 & 23 +0.270 0.030 &              \\
\hline 
0362   & 71 +0.091 0.189 & 87 +0.328 0.042 & 89 +0.248 0.040 & 91 +0.334 0.026 & 86 +0.172 0.037 &-1.166 0.048  \\
P only & 16 +0.219 0.097 & 14 +0.347 0.034 & 15 +0.231 0.027 & 16 +0.340 0.030 & 15 +0.161 0.022 &              \\
\hline 
1851   & 96 +0.002 0.209 &123 +0.374 0.036 &124 +0.383 0.029 &125 +0.324 0.035 &122 +0.155 0.053 &-1.185 0.068  \\
P only & 30 +0.170 0.112 & 30 +0.393 0.034 & 29 +0.365 0.029 & 30 +0.317 0.029 & 30 +0.144 0.049 &              \\
\hline      
1904   & 58 +0.078 0.205 & 57 +0.272 0.040 & 68 +0.304 0.036 & 67 +0.276 0.037 & 66 +0.131 0.042 &-1.579 0.033  \\
P only & 19 +0.215 0.098 & 19 +0.276 0.038 & 19 +0.292 0.039 & 19 +0.274 0.033 & 19 +0.141 0.035 &              \\
\hline      
2808   &117 +0.001 0.371 &139 +0.272 0.156 &140 +0.296 0.054 &140 +0.328 0.023 &140 +0.213 0.033 &-1.129 0.030  \\
P only & 42 +0.308 0.058 & 46 +0.384 0.041 & 46 +0.265 0.026 & 46 +0.321 0.024 & 46 +0.215 0.033 &              \\
\hline      
3201   &110 +0.125 0.286 &131 +0.341 0.049 &146 +0.299 0.038 &150 +0.306 0.039 &118 +0.082 0.037 &-1.512 0.065  \\
P only & 35 +0.273 0.166 & 32 +0.338 0.041 & 35 +0.293 0.039 & 35 +0.293 0.035 & 34 +0.079 0.034 &              \\
\hline      
4590   & 56 +0.403 0.178 & 62 +0.364 0.066 & 54 +0.423 0.057 &120 +0.280 0.042 & 18 +0.162 0.048 &-2.265 0.047  \\
P only & 19 +0.473 0.177 & 18 +0.350 0.065 & 13 +0.409 0.049 & 19 +0.272 0.042 &  8 +0.164 0.048 &              \\
\hline      
4833   & 61 +0.243 0.270 & 52 +0.338 0.173 & 71 +0.459 0.051 & 78 +0.351 0.015 & 62 +0.167 0.023 &-2.015 0.014  \\
P only & 16 +0.456 0.102 & 14 +0.508 0.078 & 16 +0.426 0.027 & 16 +0.349 0.018 & 16 +0.166 0.026 &              \\
 \hline     
5634   &  6 +0.292 0.134 &  7 +0.519 0.032 &  7 +0.295 0.031 &  7 +0.299 0.024 &  7 +0.146 0.015 &-1.867 0.050  \\
P only &  3 +0.388 0.107 &  4 +0.512 0.047 &  4 +0.296 0.045 &  4 +0.285 0.024 &  4 +0.137 0.015 &              \\
\hline      
5904   &114 +0.139 0.282 &135 +0.422 0.052 &137 +0.320 0.034 &138 +0.371 0.028 &133 +0.181 0.027 &-1.340 0.052  \\
P only & 31 +0.387 0.122 & 31 +0.418 0.053 & 31 +0.310 0.037 & 31 +0.369 0.032 & 31 +0.182 0.029 &              \\
\hline      
6093   & 63 +0.223 0.217 & 70 +0.458 0.061 & 79 +0.354 0.031 & 82 +0.356 0.016 & 68 +0.189 0.029 &-1.792 0.023  \\
P only & 18 +0.405 0.111 & 17 +0.461 0.048 & 17 +0.351 0.037 & 18 +0.356 0.016 & 18 +0.187 0.028 &              \\
\hline      
6121   & 88 +0.218 0.118 &104 +0.545 0.045 &104 +0.538 0.039 &104 +0.415 0.035 &104 +0.281 0.033 &-1.168 0.046  \\
P only & 26 +0.313 0.079 & 26 +0.547 0.050 & 26 +0.519 0.039 & 26 +0.412 0.031 & 26 +0.277 0.034 &              \\
\hline      
6171   & 30 +0.172 0.187 & 33 +0.522 0.044 & 33 +0.524 0.049 & 33 +0.415 0.032 & 33 +0.174 0.038 &-1.033 0.064  \\
P only & 10 +0.309 0.075 & 10 +0.533 0.028 & 10 +0.528 0.056 & 10 +0.425 0.031 & 10 +0.170 0.032 &              \\
\hline      
6218   & 75 +0.281 0.275 & 81 +0.537 0.037 & 81 +0.356 0.044 & 81 +0.420 0.038 & 80 +0.251 0.024 &-1.330 0.042  \\
P only & 18 +0.525 0.137 & 18 +0.540 0.032 & 18 +0.316 0.034 & 18 +0.410 0.047 & 18 +0.254 0.025 &              \\
\hline      
6254   &109 +0.271 0.212 &129 +0.481 0.065 &144 +0.314 0.057 &152 +0.342 0.037 &130 +0.176 0.043 &-1.575 0.059  \\
P only & 33 +0.401 0.131 & 32 +0.519 0.067 & 31 +0.294 0.051 & 33 +0.328 0.037 & 33 +0.166 0.045 &              \\
\hline      
6388   &183 -0.071 0.229 &184 +0.220 0.052 &184 +0.317 0.063 &185 +0.067 0.046 &185 +0.275 0.096 &-0.480 0.045  \\
P only & 53 +0.103 0.145 & 53 +0.233 0.053 & 53 +0.284 0.048 & 54 +0.053 0.050 & 54 +0.258 0.105 &              \\
\hline      
6397   & 19 +0.274 0.085 & 97 +0.454 0.041 & 49 +0.337 0.037 &147 +0.290 0.035 & 38 +0.177 0.037 &-1.988 0.044  \\
P only &  4 +0.332 0.069 &  4 +0.467 0.040 &  4 +0.326 0.037 &  4 +0.267 0.037 &  4 +0.164 0.025 &              \\
\hline      
6441   & 29 +0.040 0.180 & 29 +0.370 0.130 & 30 +0.400 0.180 & 29 +0.170 0.180 & 27 +0.320 0.180 &-0.348 0.090  \\
P only & 11 +0.100 0.100 & 11 +0.410 0.130 & 11 +0.360 0.210 & 11 +0.110 0.170 & 10 +0.250 0.150 &              \\
\hline      
6535   & 27 +0.602 0.190 & 24 +0.479 0.051 &  6 +0.418 0.035 &  7 +0.312 0.018 &  7 +0.184 0.036 &-1.952 0.036  \\
P only &  8 +0.699 0.178 &  7 +0.495 0.035 &  7 +0.428 0.048 &  8 +0.339 0.043 &  6 +0.186 0.045 &              \\
\hline      
6715   & 76 +0.070 0.344 & 75 +0.344 0.170 & 77 +0.414 0.102 & 77 +0.363 0.107 & 77 +0.166 0.137 &-1.505 0.164  \\
P only & 21 +0.359 0.078 & 21 +0.421 0.102 & 21 +0.385 0.081 & 21 +0.307 0.083 & 21 +0.095 0.099 &              \\
\hline      
6752   &107 +0.242 0.245 &125 +0.387 0.107 &137 +0.476 0.062 &130 +0.397 0.033 &134 +0.181 0.039 &-1.555 0.051  \\
P only & 26 +0.452 0.163 & 25 +0.457 0.058 & 26 +0.434 0.064 & 26 +0.395 0.027 & 24 +0.174 0.035 &              \\
\hline      
6809   &113 +0.251 0.191 &129 +0.483 0.061 &146 +0.380 0.044 &152 +0.363 0.042 &133 +0.147 0.056 &-1.934 0.063  \\
P only & 17 +0.380 0.090 & 17 +0.523 0.041 & 17 +0.377 0.041 & 17 +0.360 0.052 & 17 +0.140 0.050 &              \\
\hline      
6838   & 43 +0.364 0.112 & 51 +0.496 0.043 & 51 +0.391 0.051 & 51 +0.312 0.050 & 51 +0.366 0.070 &-0.832 0.061  \\
P only & 12 +0.403 0.109 & 12 +0.499 0.042 & 12 +0.391 0.056 & 12 +0.274 0.050 & 12 +0.342 0.067 &              \\
\hline      
7078   & 45 +0.308 0.187 & 55 +0.420 0.175 & 59 +0.472 0.093 & 82 +0.273 0.054 & 36 +0.219 0.040 &-2.320 0.057  \\
P only & 13 +0.438 0.125 & 13 +0.542 0.107 & 10 +0.395 0.081 & 13 +0.254 0.053 & 12 +0.213 0.029 &              \\
\hline      
7099   & 35 +0.376 0.264 & 33 +0.505 0.044 & 22 +0.367 0.058 & 68 +0.323 0.051 & 24 +0.240 0.032 &-2.344 0.049  \\
P only & 12 +0.507 0.115 & 10 +0.511 0.035 &  5 +0.298 0.062 & 12 +0.318 0.045 &  8 +0.234 0.031 &              \\
\hline      
Ter8   &  6 +0.394 0.045 &  6 +0.470 0.092 &  6 +0.248 0.095 &  6 +0.190 0.038 &  6 +0.050 0.057 &-2.271 0.079  \\
P only &  6 +0.394 0.045 &  6 +0.470 0.092 &  6 +0.248 0.095 &  6 +0.190 0.038 &  6 +0.050 0.057 &              \\
     
\hline
\end{tabular}

\begin{flushleft}
Notes: for each GC, the second row lists the average abundances computed using
only the P stars with primordial composition; n is the number of stars used in
the averages.
\end{flushleft}

\label{t:tableA2}

\end{table*}

We added other 11 GCs studied by the same group in individual papers to the
sample in Carretta (2026). Average abundances of oxygen and other
$\alpha-$elements are listed in Table~\ref{t:tableA2} in the first row for each
GC, together with the cluster metallicity [Fe/H] on the metallicity scale by
Carretta et al. (2009b) from high resolution UVES spectra. We stress that this
ensemble of 27 GCs represents a very homogeneous dataset, being analysed with
identical methodology. We employed the same scale of atmospheric parameters, the
same line list, the same method for measurements of equivalent widths, the same
reference solar abundances. All the stars (more than 2600 with Fe and at least a
measured species of $\alpha-$elements) have membership certified by radial
velocity and abundance. References for all the papers where abundances of
individual stars can be retrieved (usually in electronic form at CDS) are given
in Table~\ref{t:tableA2ref}.

We adopted the classification given in Massari et al. (2019), with a few
amendments as in Carretta (2026), for the birth place of GCs. In our sample,
eight GCs are in situ clusters, born in the disc or bulge of the Milky Way. The
other objects are though to have formed in former external systems and afterward
brought in the main Galaxy from the accretion events linked to Gaia-Enceladus,
Sequoia, the Helmi streams, Sagittarius, or the group of low orbital energy GCs
of still uncertain origin.

We used field stars of the Milky Way as a reference baseline. We employed the
sample of stars with good parallaxes by Gratton et al. (2003) from which we also
adopt our solar reference abundances. The field stars cover the same metallicity
range spanned by the GC sample and the abundances are derived from the same line
list as used for GC stars. Another reference consists in 27  stars in the
nucleus of the Sgr dSph from Carretta et al. (2010c).
Although limited in size, this sample is useful to evaluate the  behaviour of
the HEx ratio in a substructure accreted in the Milky Way. From the comparison
of the abundances, Carretta et al. (2010c) showed that the stars in the Sgr core
can be distinct from those of the GC NGC~6715. While presently embedded in the
Sgr core, this GC probably formed separately and  plunged in the core of Sgr due
to dynamical friction only at later times. The different chemistry of core stars
and GC is clearly highlighted by proton-capture elements (Carretta et al.
2010c), but it seems also shared by the ratio of hydrostatic to explosive
$\alpha-$elements (see below).

\begin{figure}[t]
\centering
\includegraphics[scale=0.40]{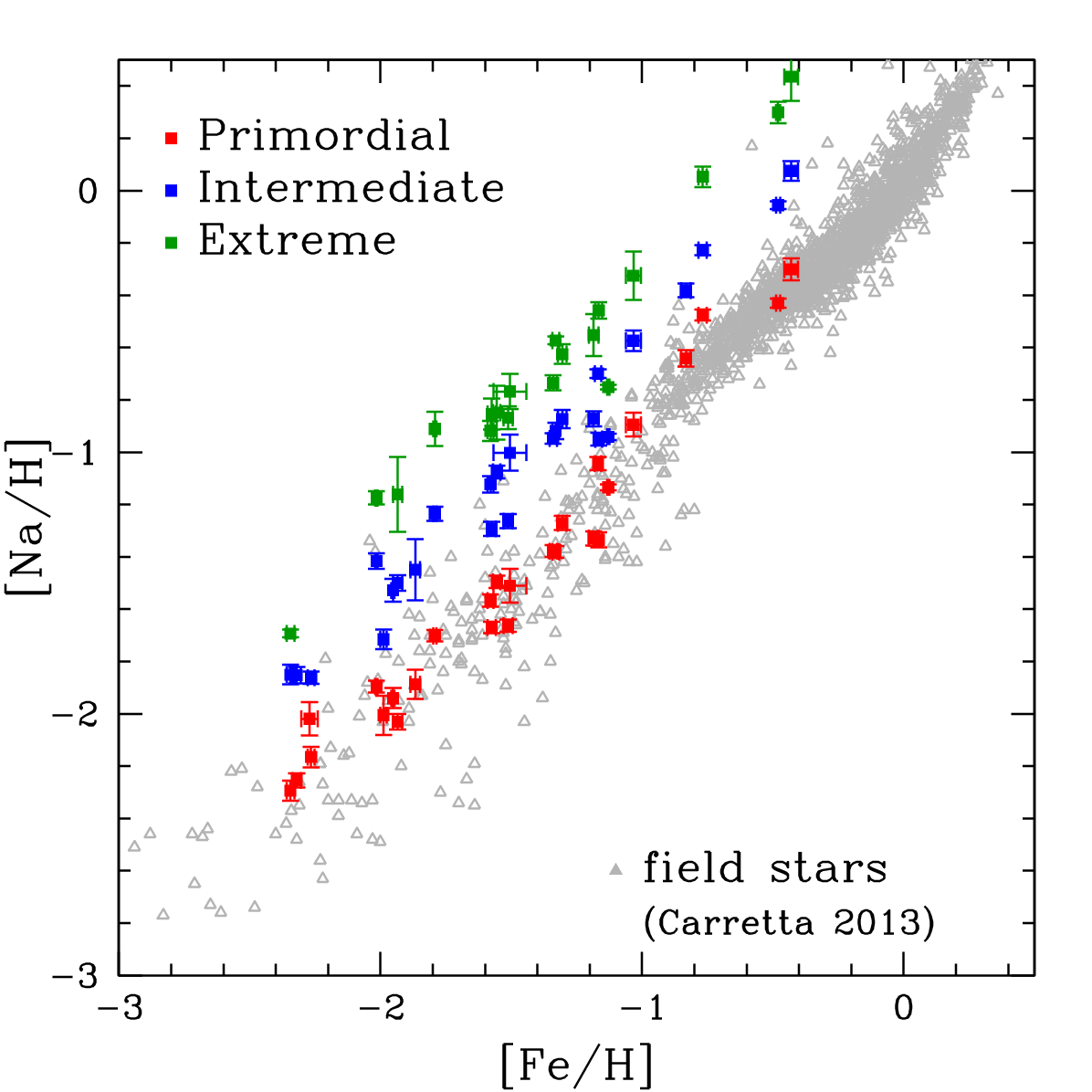}
\caption{Average abundance ratio [Na/H] as a function of metallicity for GC 
stars with primordial (P, red  squares), intermediate (I, in blue), and extreme
(E, green) composition in our  GCs, superimposed to a reference sample of
unpolluted field stars from Carretta (2013: grey open triangles).}
\label{f:nahmed}
\end{figure}

The impact of the properties of MPs is a potential problem in deriving the
correct or unbiased HEx ratio for GCs. Abundance variations due to
proton-capture reactions in H-burning at high temperature affect in particular
the hydrostatic species O and Mg, and only to a smaller extent explosive
$\alpha-$elements like e.g. Si and Ca (see e.g. section 1 and Carretta 2026).
This potential bias can be efficiently solved by using only the unpolluted
stellar component in GCs, that is the fraction of stars that still maintain the
original chemical composition, before the polluters enriched the proto-cluster
in ejecta processed by proton-capture reactions.

To this purpose, we adopt the criteria given in Carretta et al. (2009c) for the
classification of MPs in GCs according to the Na-O anti-correlation. The
unpolluted component with primordial (P) composition includes all stars with Na
abundance between the minimum [Na/Fe] observed in each GC and  [Na/Fe]$_{\rm
min}+4\sigma$. The polluted stars are separated into fractions with intermediate
(I) and extreme (E) composition according to their position along the Na-O
anti-correlation ([O/Na]$>-0.9$ and [O/Na]$<-0.9$ dex respectively). Concerning
the present study, what is relevant is that the P stars bear the composition of
unpolluted field stars of similar metallicity, inherited essentially from SNe
nucleosynthesis, as shown in Fig.~\ref{f:nahmed}. In this figure we plot the
average [Na/H] ratios for the P, I, and E components in our sample of GCs
superimposed to a reference sample of field stars from Carretta (2013). The
P stars well matches the locus of the unpolluted stars. 
The good correlation between abundances of O and Mg in MPs (p-value $<1.0\times
10^{-6}$ in our sample) ensures that the Mg abundances used in the HEx ratio
reliably trace the birth-gas composition. We prefer to use the Na-O
anti-correlation as a tool to select the P stars because it is a signature
present in the overwhelming majority of GCs, whereas significant Mg variations
are biased towards massive and metal-poor GCs only (see Section 1). Average
abundances of the P component are listed in the second row in 
Table~\ref{t:tableA2} for each GC.

\section{The HEx ratio}

\begin{figure}
\centering
\includegraphics[scale=0.40]{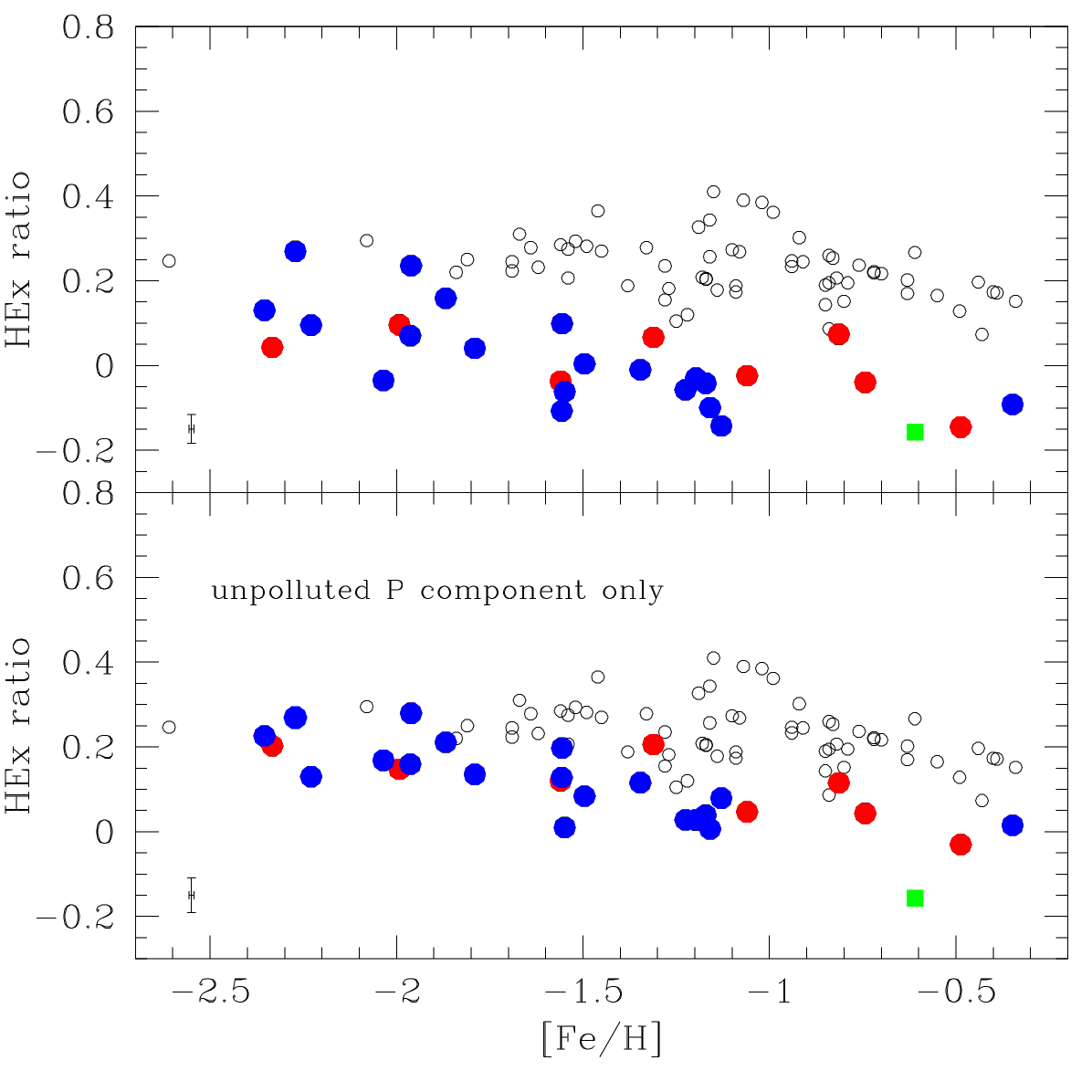}
\caption{The ratio of the average of hydrostatic $\alpha-$elements O and Mg to
the average of explosive $\alpha-$elements Si, Ca, and Ti (HEx ratio) for our
sample of GCs. In the lower panel the ratio is computed using only stars of the
unpolluted P component in each GC (see text). Blue and red symbols are for
accreted and in situ GCs, respectively, grey empty circles are field stars from
Gratton et al. (2003). The green square is for field stars in the core of Sgr
dSph. A typical error is shown in both panels (average from all GCs).}
\label{f:hex1a}
\end{figure}

Derived HEx ratios of hydrostatic to explosive $\alpha-$elements is shown in
Fig.~\ref{f:hex1a} as a function of metallicity for our sample of GCs. The ratio
is computed from the mean ([O/Fe]+[Mg/Fe])/2 and ([Si/Fe]+[Ca/Fe]+[Ti/Fe])/3 for
hydrostatic and explosive species, respectively. Red colour is used for in situ
GCs and blue colour indicates accreted GCs. Empty grey circles are field
galactic stars used as a comparison. The green square is the value for the stars
in the nucleus of Sgr. 

In the upper panel, the averages are done using all stars in GCs, regardless of
the P, I, and E class, whereas in the lower panel we use only the unpolluted
P component. The two panels allow to discern at a glance the effect of the MP
phenomenon. In GCs, O can be depleted by as much as 1 dex by  proton-capture
reactions at moderate temperature (20-40 MK, see e.g. Gratton et al. 2019),
whereas the depletion in Mg is usually much smaller, requiring higher
temperature for a significant variation. Compared to the reference field stars
the HEx  ratios in the upper panel are lower due to the large depletions in O
and the moderate depletions in Mg in the polluted components I and E of GCs.
When only the unpolluted component is used (lower panel) the HEx ratios are
higher, closer to the locus of field stars.  

In both panels the ratio is decreasing with increasing [Fe/H], irrespective of
the origin (in situ or accreted) of GCs, in good agreement with results by HN25
from APOGEE. The HEx ratio for field stars in the Sgr nucleus is located below
the plane defined by the GCs, in particular in the lower panel. Again, this is
consistent with the overall lower ratio found by HN25 for satellite galaxies when
compared to GCs. Field stars (empty circles) seems to show a pattern with a knee
at [Fe/H]=-1 dex reminiscent of the classical plateau plus decrease displayed by
individual $\alpha-$elements.

We note that Horta et al. (2020) claim that a difference exists between the mean
level of [Si/Fe] for accreted and in situ subgroups of GCs at metallicity higher
then about [Fe/H]=-1.5. However, in both HN25 and in the present study there is
no dependence on the origin of the various groups of GCs. In the HEx ratio the
weight given to Si is decreased by ``diluting" the relevance of the [Si/Fe]
ratio with four other $\alpha-$elements. These findings seem to corroborate the
possible presence of problems in the analysis of Si in APOGEE, as suggested in
Carretta (2026).

\section{Comparison with APOGEE results}

Although the trend with [Fe/H] of the HEx ratio is broadly consistent with the 
results by HN25 and we agree that in our homogeneous dataset the locus of 
accreted and in situ GCs cannot be distinguished,  we found a few relevant
differencies with APOGEE results. First, HN25 claim that NGC~7078
(M~15) is an outlier, with a low HEx ratio ($\sim 0.1$) at [Fe/H]$\sim -2.45$.
In our data M~15, with a HEx ratio 0.203 is well inside the region populated by
the other metal-poor  GCs (e.g. NGC~4590, NGC~7099, Ter~8).

\begin{figure}
\centering
\includegraphics[scale=0.40]{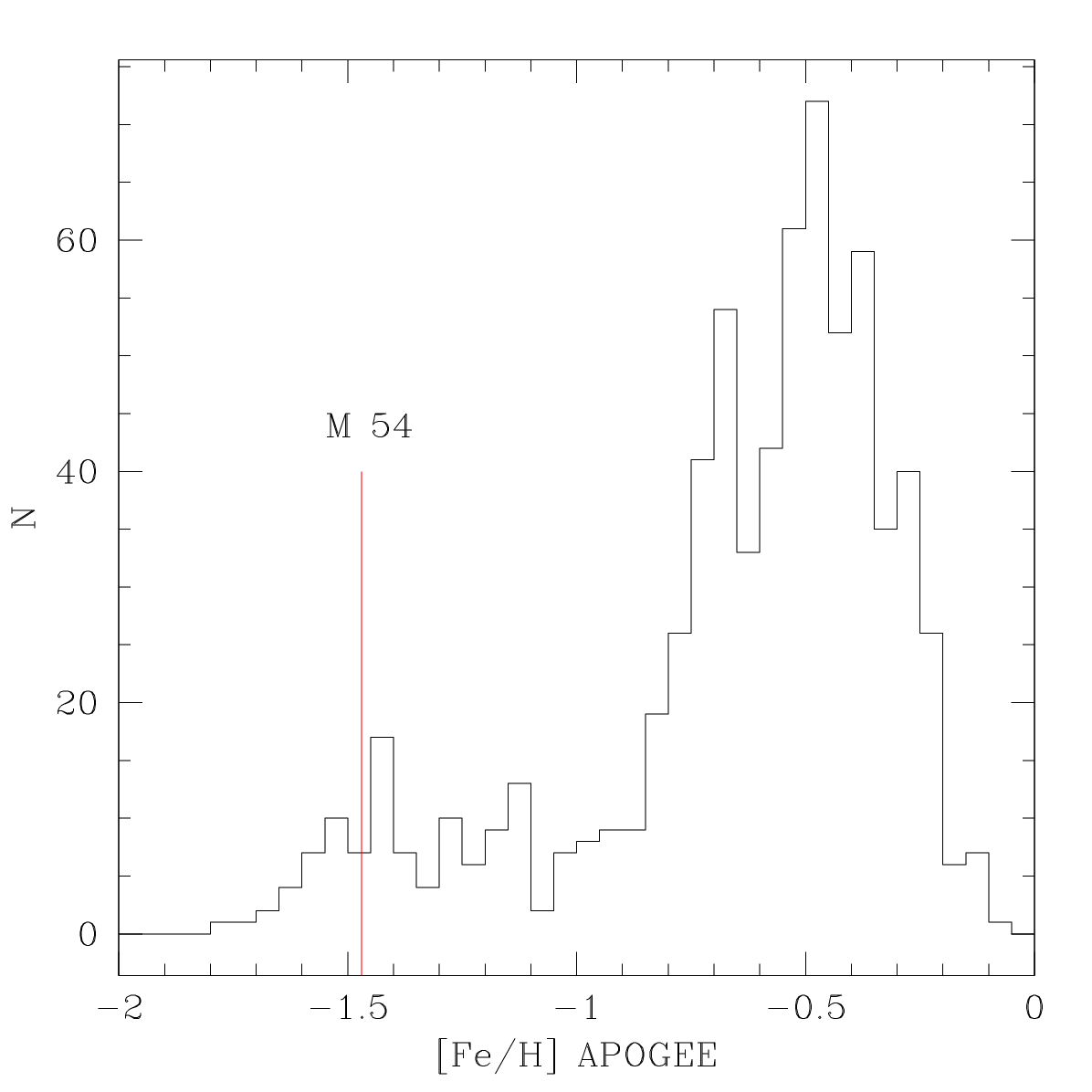}
\caption{Distribution of metallicities for stars labelled as candidate members of
NGC6715 in Schiavon et al. (2024), after the selection cuts from Horta and Ness
(2025). The mean metallicity from APOGEE (M\'esz\'aros et al. 2020) for M~54 is
indicated by the vertical red line.}
\label{f:histoHN}
\end{figure}

Second, HN25 find for NGC~6715 (M~54) an HEx ratio very close
to the value derived for the core of the Sgr galaxy, at [Fe/H]$\sim -0.65$ dex.
This is highly surprising, since the average metallicity of M~54 is
-1.559\footnote{This value is different from the one in Table~\ref{t:tableA2}
because is the average from the more large sample of GIRAFFE stars.} 
($\sigma=0.189$ dex, 76 stars) according to Carretta et al. (2010c), who also find a 
mean [Fe/H]$=-0.622$ dex from the 27 stars in the Sgr nucleus surrounding M~54. 
Furthermore, the APOGEE results in M\'esz\'aros et al. (2020) give a mean
metallicity [Fe/H]$=-1.47$ ($\sigma=0.15$ dex) for M~54, again inconsistent with
the value in HN25. 
To better understand the origin of the discrepancy, we cross-identified the 76
bona fide members of M~54 for radial velocity and abundance in Carretta et al.
(2010c) with APOGEE DR17. We found 16 matches, with 6 stars flagged
ASPFLAG=0 denoting reliable abundances. Three out of six are unpolluted stars,
according to Carretta et al. (2010c). From them we derived [Fe/H]=-1.482 dex
and HEx=+0.140, in excellent agreement with the HEx sequence of GCs of similar
metallicity. As a check, we next followed the procedure of HN25 by extracting
from the value-added catalogue by Schiavon et al. (2024) all entries for
NGC6715. We then applied all the selection criteria listed in Section 2 of HN25.
The distribution of the metallicities from this selection is shown in
Fig.~\ref{f:histoHN}, where the bulk is at [Fe/H] around -0.6 dex, consistent
with the average metallicity for stars in the Sgr nucleus found in Carretta et
al. (2010c). In HN25, both M~54 and Sgr seem to be located at this mean value of
[Fe/H]. We hypothesise that probably
the selection for the label NGC6715 provided mostly stars of the Sgr core (609
stars), in which the genuine globular cluster M~54 is immersed (the peak at -1.5
dex, 98 stars). The  close values shown in HN25 are likely the
result of this  confusion. The genuine cluster M~54, whose stars show an
extended Na-O anti-correlation (signature of a true GC), is much more metal-poor
(Carretta et al. 2010c, M\'esz\'aros et al. 2020). At this metallicity, the
hydrostatic and explosive $\alpha-$elements of its stellar population give a HEx
ratio (+0.128 from our analysis) that places M~54 in the midst of the relation
defined by GCs, being well distinct from the value (lower HEx ratio and higher
metallicity) derived from field stars in the Sgr core (see
Fig.~\ref{f:hex1a}). Moreover, in our dataset Terzan 8 (HEx=0.269), 
another GC associated to the Sgr dSph, is well located on the sequence of GCs
for its metallicity and well separated from the Sgr field location.

\section{Discussion and conclusions}

\begin{figure}
\centering
\includegraphics[scale=0.40]{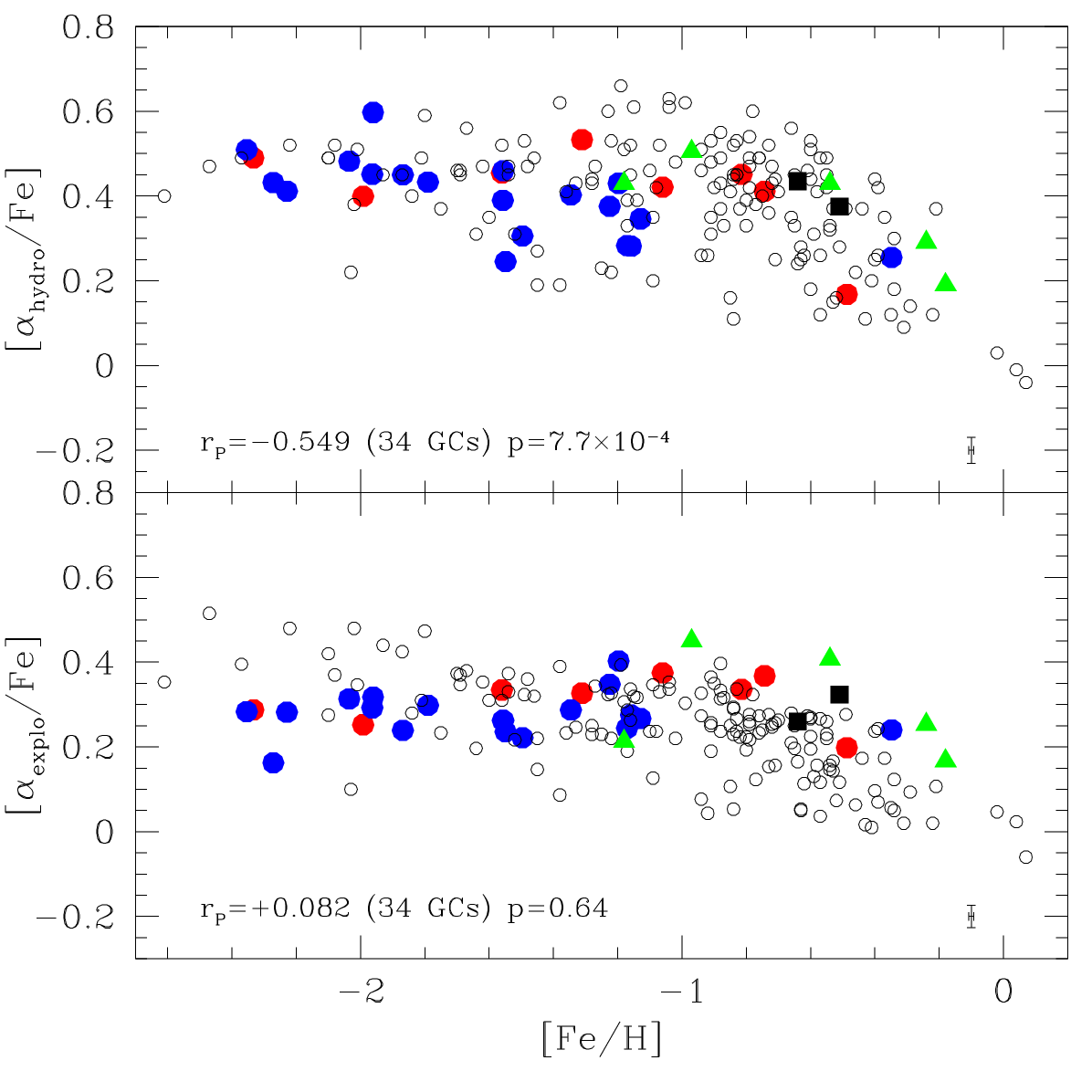}
\caption{Abundance ratios of the hydrostatic (upper panel) and explosive (lower
anel) $\alpha-$elements in the GCs of our extended sample, Blue and red points
indicate accreted and in situ GCs. Black squares are disk GCs and green
triangles are bulge GCs from literature. The Pearson's correlation coefficient and
two-tailed probability are listed. Only Mg is used to represent the hydrostatic
species in field stars.}
\label{f:hydroexplo}
\end{figure}

A large survey of $\alpha-$elements in GCs, recently completed, is exploited 
together with previous studies to assemble an extremely homogeneous dataset for
abundances of hydrostatic and explosive elements in 27 GCs. From this set we
computed the HEx ratio (Carlin et al. 2018) of hydrostatic (O, Mg) to explosive (Si, Ca, Ti)
elements, using the primordial P component in GCs to disentangle the original
ratios from the effects of MPs that alter predominantly the abundances of O,
but also those of Mg, Si, and sometimes Ca (see e.g. Carretta 2026).

In general, despite a few discrepant cases, we confirm with high resolution
optical spectroscopy the main results obtained by HN25 using the near-infrared
spectroscopic survey APOGEE. In the plane mapping the HEx ratio as a function of
metallicity, the GCs cannot be distinguished by their in situ or accreted
origin. In this plane, the sequence of GCs born in the main body of the Milky Way
is superimposed to the sequence of GCs though to be formed in external systems.
Both sequences lie about 0.1 dex above the locus occupied by satellite galaxies.
In our study, we have only one comparison represented by stars in the Sgr dSph
core, for reasons of homogeneity in the abundance analysis. However, this
finding is supported by the much larger sample of galaxies and halo
substructures shown in HN25.

We also confirm that the HEx ratios of GCs decrease as the metallicity
increases, as shown in HN25. The observed gradient with [Fe/H] can be explained
following the two scenarios first discussed by McWilliam et al. (2013) to
account for the variation of the [Mg/Ca] ratio over the metallicity range of the
Sgr dSph field stars. The HEx ratio measures the contribution to
$\alpha-$elements by hydrostatic burning in massive stars compared to the yield
from all the stars exploding as SNe events. The two scenarios affect
respectively the numerator and the denominator of the HEx ratio.

\begin{figure}
\centering
\includegraphics[scale=0.40]{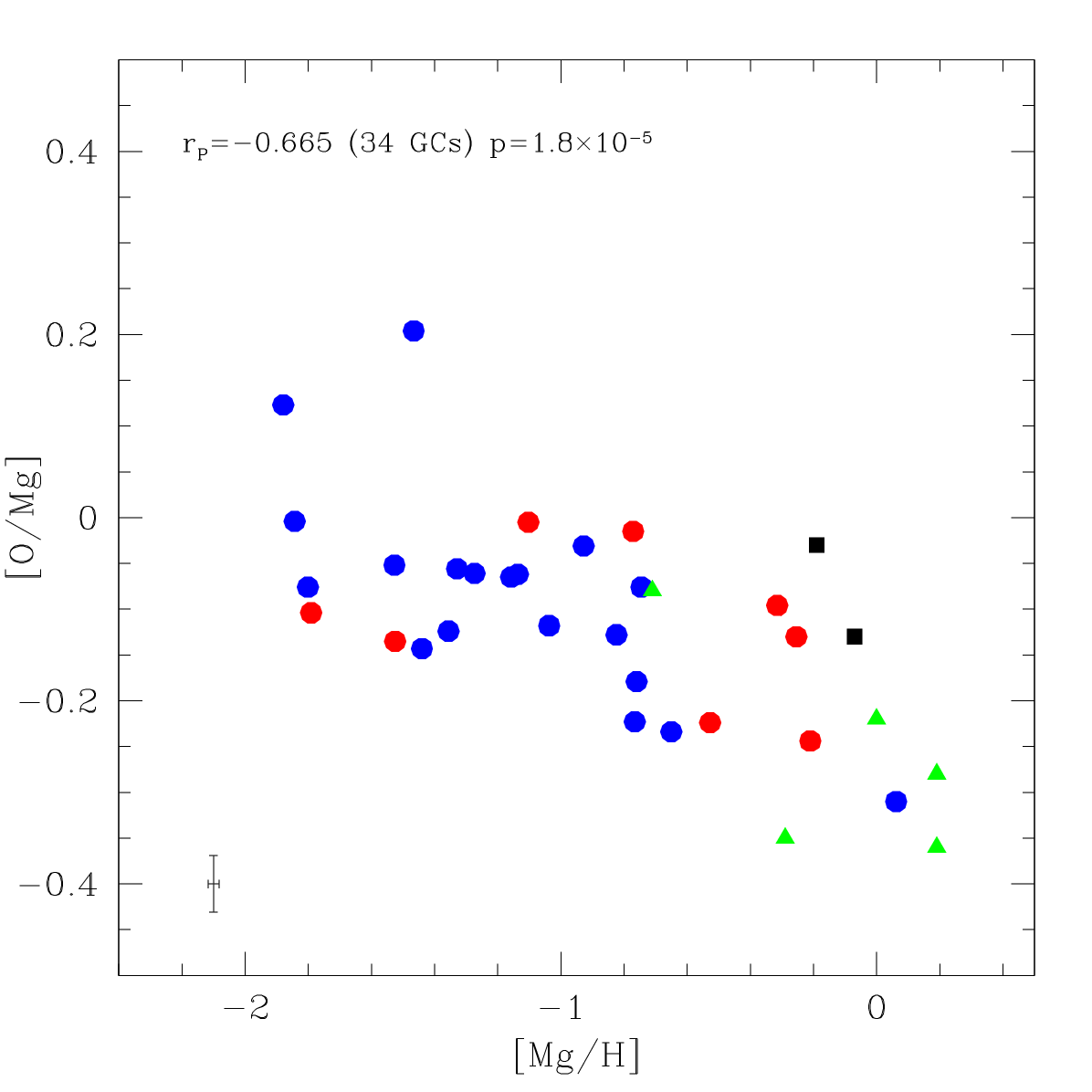}
\caption{Abundance ratio [O/Mg] as a function of [Mg/H] for the extended sample
of GCs. Symbols are as in Fig.~\ref{f:hydroexplo}.}
\label{f:robaH}
\end{figure}

A lower HEx ratio may imply a smaller contribution to the enrichment by the most
massive stars ($M > 8 $M$_\odot$) among the SN II progenitors, that is a top
light (bottom heavy) IMF like that typical of low mass satellite galaxies of the
Milky Way. This scenario explains the position of dwarf galaxies in the
HEx ratio - [Fe/H] plane, sistematically located at lower HEx ratios with
respect to GCs at any metallicity (see HN25, and also the value for Sgr in 
Fig.~\ref{f:hex1a} of the present paper). 

Alternatively, a lower HEx ratio may result from changing the factors that
control the denominator, for example an excess of type Ia SNe. An increased 
contribution to the nucleosynthesis as metallicity increases will raise both
[Fe/H] and the content of Si, Ca, and  Ti, whereas the production of hydrostatic
elements is expected to be negligibly low (e.g. Nomoto et al. 1984). 
This is the explanation preferred by HN25 to account for the trend with metallicity
observed for the HEx ratio in GCs. 

On the other hand, in a single dSph, Sgr, Carlin et al. (2018) found that the
abundance ratios of hydrostatic elements were located much further below the MW
stars used as reference that the abundance ratios of elements produced in
explosive nucleosynthesis, suggesting a IMF lacking the most massive SN II
progenitors.

It seems to be even more difficult to use a delayed SNe Ia mechanism to explain
the decreasing HEx ratio with metallicity for GCs that are originated in
different environments. The chemodynamics and the age-metallicity relations of
GCs in Fig.~\ref{f:hex1a} indicates that they come from five or six galaxies
(Massari et al. 2019), i.e. the Milky Way disc and bulge, Gaia-Enceladus,
Sequoia, Sgr, and the progenitors of the Helmi Stream and low-energy GCs, later
injected in the Milky Way following the accretion events. It would be hard to
imagine that the SNe Ia contribution worked to place exactly on the same
relation of declining HEx ratio GCs formed in systems with different global
masses, star formation rates, and thus different timescales for the prompt and
delayed enrichment from SNe. 

The two alternative scenarios can be tested by looking separately at the trends 
with the metallicity of each of the two terms of the HEx ratio, 
[$\alpha_{hydro}$/Fe] and [$\alpha_{explo}$/Fe],  as done in
Fig.~\ref{f:hydroexplo}. For this figure we employed again only the average
values derived from the primordial P component in each GC, to remove the effect
of MPs on the ratios. We adopt only Mg as representative of hydrostatic species
in field stars to increase the sample size. To better sample the critical region
at  high metallicity ([Fe/H]$\gtrsim -1$ dex) where the bulk of SNe Ia is
expected to manifest, we added to our sample GCs from literature. We included
five bulge  GCs (green triangles in Fig.~\ref{f:hydroexplo}) and two disc GCs
(black squares), all with abundances derived from high resolution optical
spectra. The added GCs are NGC~6553, NGC~6528, NGC~6440 (Mu\~{n}oz et al. 2020,
2018, 2017), NGC~6723 (Crestani et al. 2019), NGC~6522 (Barbuy et al. 2021),
NGC~5927 (Mura-Guzm\'{a}n et al. 2018), and NGC~6366 (Puls et al. 2018). All
abundances were corrected to our solar abundance scale. The stars analysed in
these GCs are often less than 10, hence we could not safely apply the
classification by Carretta et al. (2009c) to extract the unpolluted component.
However, in these metal-rich GCs the Na-O anti-correlation has an almost
vertical trend (e.g. Mu\~{n}oz et al. 2020), with no significant spread in O,
and the Mg-Al anti-correlation, with significant Mg depletions, is not observed
in metal-rich GCs (Carretta et al. 2009a, M\'esz\'aros et al. 2020).

We applied a linear regression to the extended sample of GCs in
Fig.~\ref{f:hydroexplo}. We list in the panels the Pearson's correlation
coefficient and  the two-tailed probability testing the null hypothesis that the
observed value  comes from a population in which the true correlation is zero.
In the lower panel of Fig.~\ref{f:hydroexplo} the (weak) correlation  of the
ratio  [$\alpha_{\rm explo}$/Fe] as a function of metallicity is not
statistically significant, meaning that the level of explosive $\alpha-$elements
is almost constant in all GCs over the whole metallicity range. Instead, the 
effect due to the time delay between core-collapse and SNe Ia as detailed by
Tinsley (1979) is clearly visible in the reference field stars after about
[Fe/H]=-1 dex, the region where stars enter the high-Ia regime. This
terminology, mutuated from Griffith et al. (2019), is more appropriate to
highlight that the ratio in field stars is decreasing due to the increasing
contribution of Fe from SNe Ia that effectively does lower the plateau
established by core-collapse SNe. This lowering is not apparent among GCs, with
[$\alpha_{\rm explo}$/Fe] remaining constant over an interval of about 2 dex in
[Fe/H].

\begin{figure*}
\centering
\includegraphics[scale=0.30]{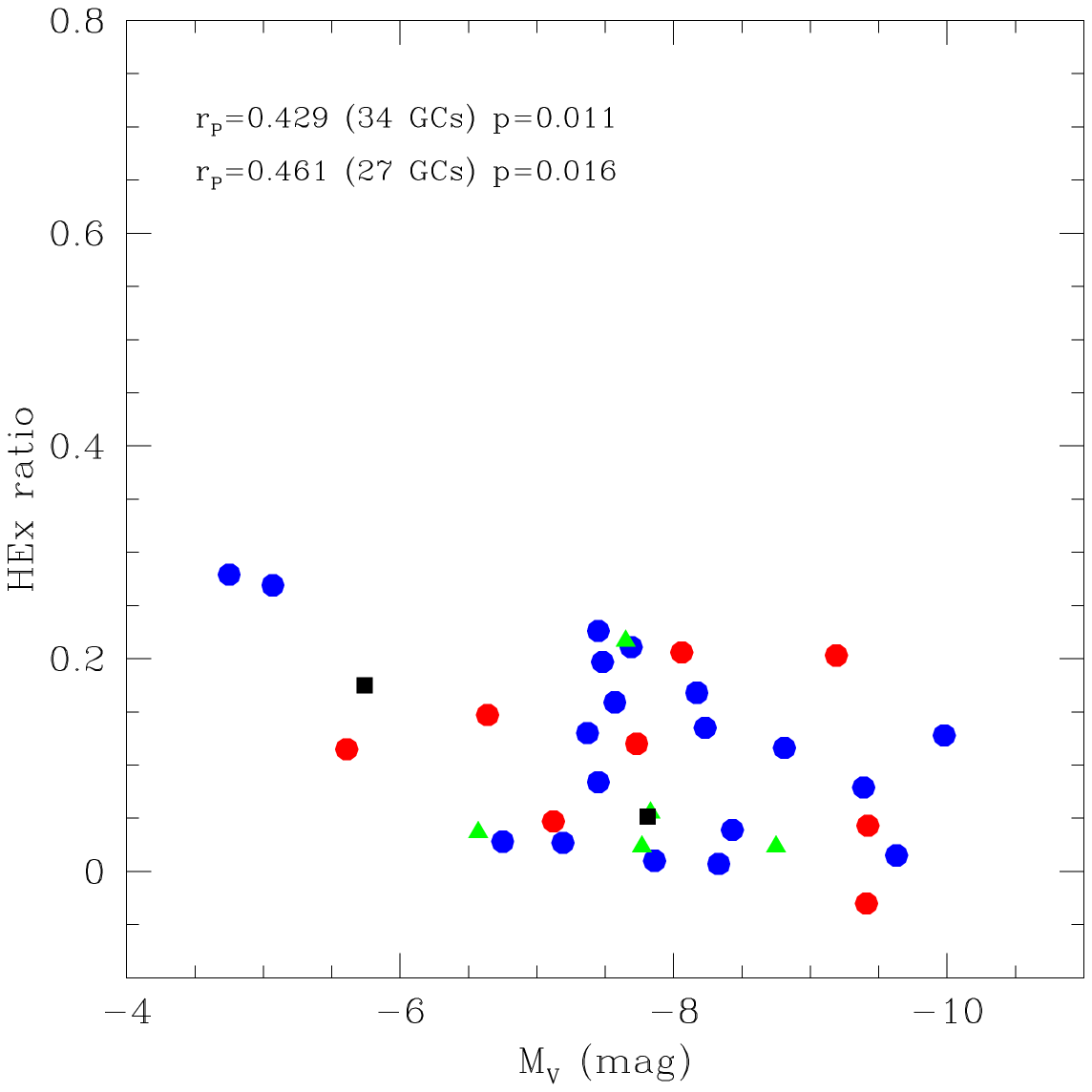}\includegraphics[scale=0.30]{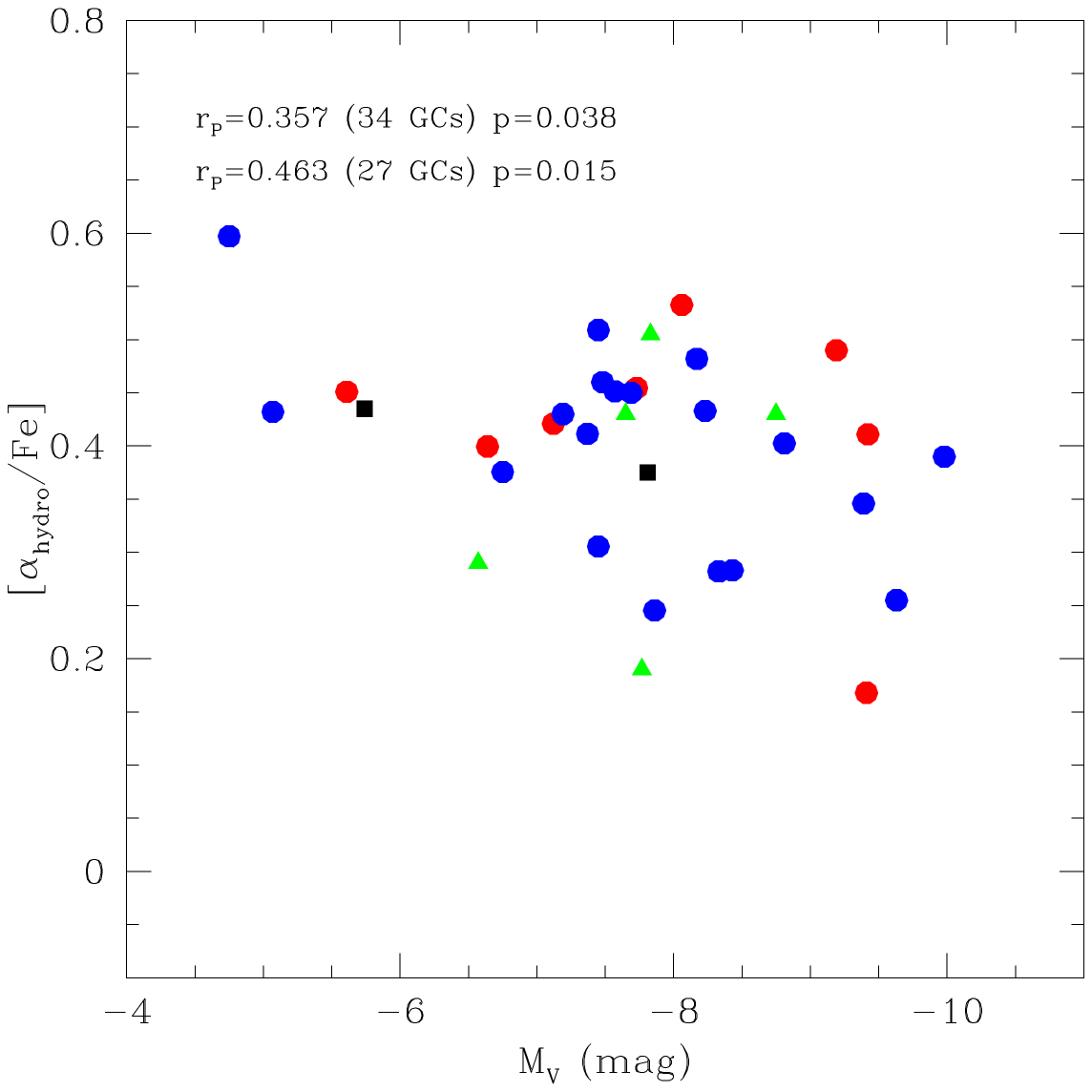}\includegraphics[scale=0.30]{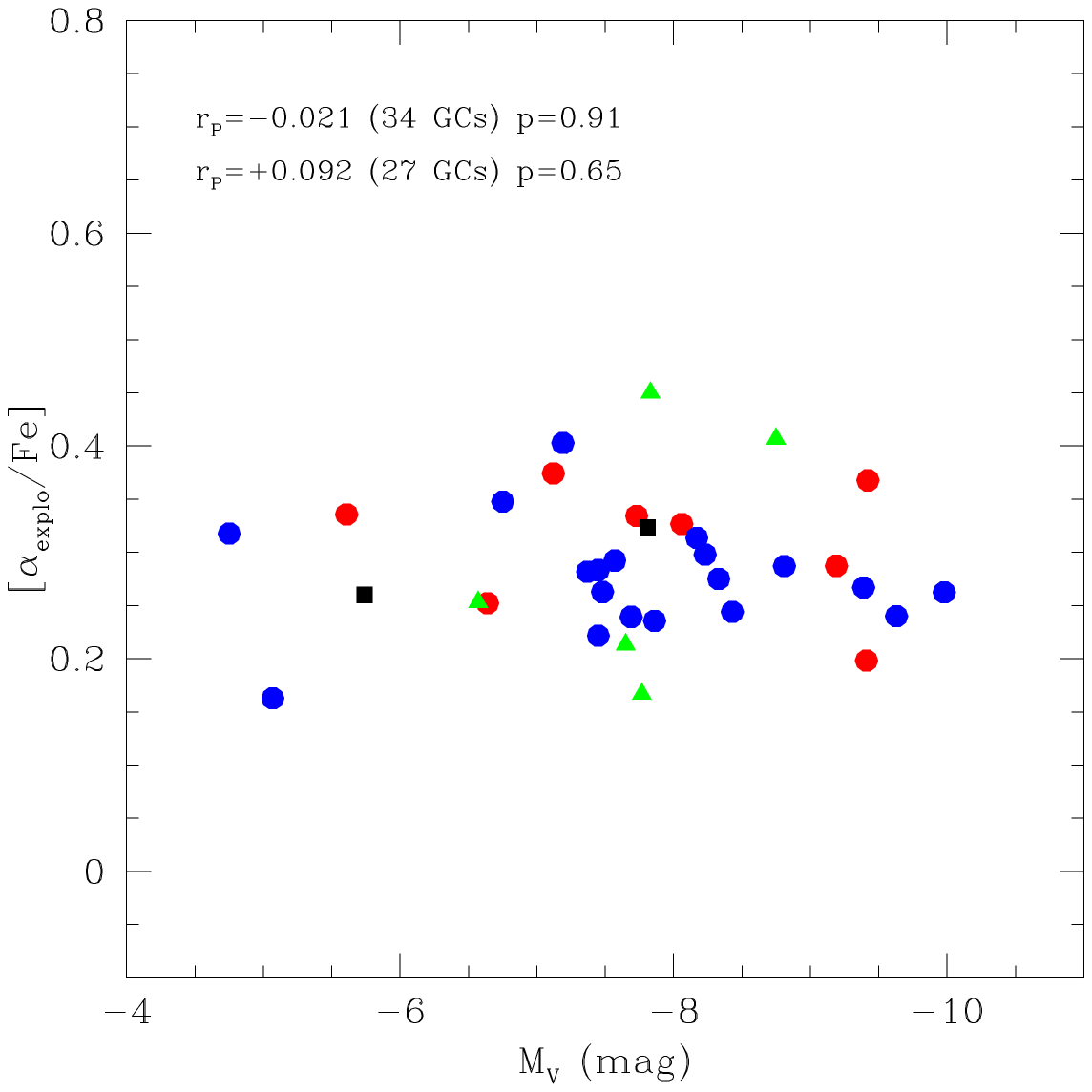}
\caption{The HEx ratio (left panel), the [$\alpha_{\rm hydro}$/Fe] (middle
panel), and the [$\alpha_{\rm explo}$/Fe] ratio (right panel) as a function of
the global absolute luminosity of GCs (Harris 2010). The results of the linear regression are
referred to the extended sample (upper row) and to our original sample (lower
row). Symbols are as in Fig.~\ref{f:hydroexplo}.}
\label{f:conMV}
\end{figure*}

What is definitely changing with regularity in GCs, already at low metallicity,
is the content of hydrostatic species (upper panel in Fig.~\ref{f:hydroexplo}).
The linear regressions listed in Fig.~\ref{f:hydroexplo} shows a
statistically significant trend with [Fe/H]. This decline is
suggesting that the impact on the HEx ratio is due to a variation of the
properties of the most massive stars (15-30 M$_\odot$). 

The irrelevance of contributions from SNe Ia can be better appreciated from
Fig.~\ref{f:robaH} where we plot the ratio [O/Mg] of the elements produced in
the hydrostatic cores of massive stars as a function of metallicity. We here
adopt the ratio [Mg/H] as reference for metallicity because Mg is produced 
almost entirely by core-collapse SNe while Fe is provided both in SNe II and Ia.
Using [Mg/H]\footnote{Our results would be the same by using [O/H] instead.} the
effect of SN Ia to the enrichment is not contained in this plot (e.g. Gratton et
al. 1996, 2000;  Furhman 1998). 

Despite O and Mg being produced by the same stars on similar timescales, the ratio
[O/Mg] is not flat, but decreases with metallicity and the relation has high
statistical significance. The expected yields of hydrostatic elements increase
rapidly with progenitor mass (e.g. Woosley and Weaver 1995) and with such a
strong sensitivity to SN mass, the IMF would impact on the abundance ratios. At
constant mass the metallicity dependence is weak (see Andrews et al. 2017),
however the IMF-averaged yields may depend on metallicity if the IMF changes as a
function of metallicity, increasing the fraction of low-mass SNe toward higher
metallicity (a more top-light IMF).
The decline of [O/Mg] with increasing [Mg/H] would be then explained
preferentially by a deficit in the highest mass stars for the more metal-rich GC
population.

There are several studies pointing to a metallicity dependent IMF for GCs. For
instance, Marks et al. (2012) found that the lower the cluster metallicity, the
more top heavy  the IMF must have been, since the high mass IMF slope is
required to be linearly correlated with metallicity in order to remove residual
gas from the proto-GC shortly after the initial star formation. The same
conclusion was reached by Wirth et al. (2022) with analytical models tailored to
investigate the observed iron spreads in GCs. They found that all GCs had IMFs
more top-heavy than the canonical one, and that a lower metallicity decreases
the high mass slope making the GC more top-heavy.
The HEx and [O/Mg] ratios, declining with increasing metallicity, support these
studies, agreeing with a scenario with an IMF that lacks more and more massive
stars as [Fe/H] increases, with no compelling need of a significant
contribution of Type Ia supernova pre-enrichment.

Finally, at odds with HN25, we find a statistically significant relation between the
HEx ratio and the total present-day mass of GCs, using as proxy the total
absolute luminosity (Fig.~\ref{f:conMV}, left panel), with
the ratio decreasing for higher mass GCs. The regression is significant using
both our extended sample (upper row label in the panel) and our original sample of GCs.
Again, this correlation is only due to the variation in the content of the
elements produced in hydrostatic burning (middle panel in Fig.~\ref{f:conMV}),
whereas the [$\alpha_{\rm explo}$/Fe] ratio remains constant (right panel).

\begin{figure}
\centering
\includegraphics[scale=0.40]{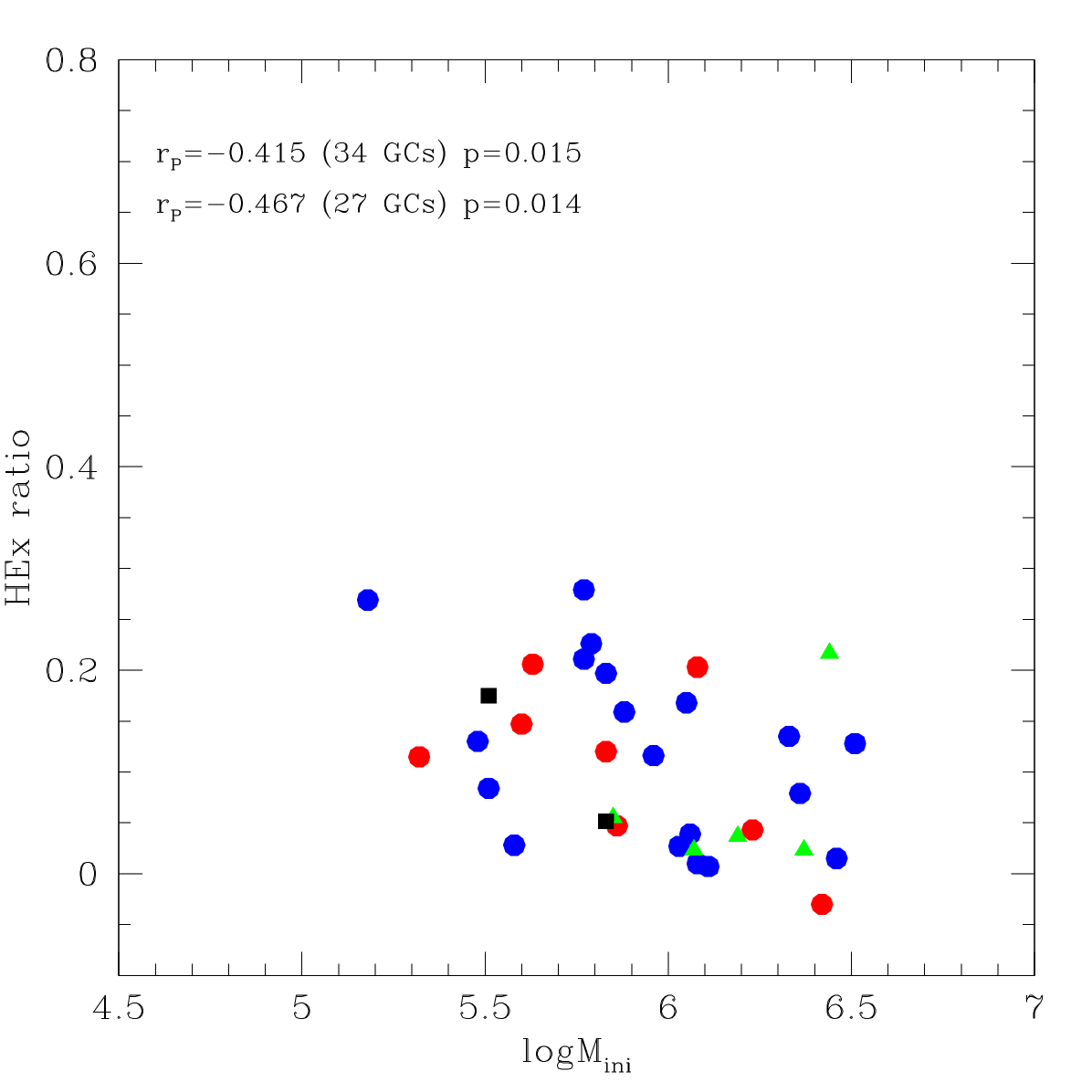}
\caption{The HEx ratio as a function of the initial GC mass from Baumgardt et al
(2019). Symbols are as in Fig.~\ref{f:hydroexplo}.}
\label{f:conMVmini}
\end{figure}

Present day masses could not be the best choice for objects like GCs,  formed in
a variety of different environments and subjected to the effects of  dynamical
evolution for almost an Hubble time. However, it seems that the relation between
the HEx ratio and the mass is imprinted since the epoch of  the GC formation. 
Indeed, we found a statistically significant decrease of the HEx ratio also as a
function of increasing initial GC masses (estimated by Baumgardt et al. 2019),
as shown in Fig.~\ref{f:conMVmini}, again in contrast with results from HN25,
who do not find any correlation with GC mass. 

We speculate that the decline with total mass could be a reflection of the weak 
correlation between mass and metallicity among GCs, since the most massive
objects are preferentially closer to the central regions of the Galaxy, where
it is more probable to find metal-rich objects (see e.g. the univariate
correlations in figure 5 of Baumgardt et al. 2019).

To summarise, all the signatures we found in the present work (the declining
ratio [O/Mg] as a function of [Mg/H], the HEx ratio decreasing with increasing
metallicity, and with initial and present day mass of GCs)   would seem to
support a coherent picture where a IMF changing with metallicity may explain the
observed correlations. Even if the IMF in the high mass regime can never be
directly observed in present day GCs, the content of different $\alpha-$elements
incorporated in the stars we can access today can be  efficiently used to probe
the ephemeral first phases of GC evolution.

\begin{acknowledgements}
This research has made large use of the SIMBAD database (in particular  Vizier),
operated at CDS, Strasbourg, France, of the NASA's Astrophysical Data System,
and TOPCAT (Taylor 2005). I especially thanks Angela Bragaglia and Donatella
Romano for several valuable suggestions. I also aknowledge funding from Bando
Astrofisica Fondamentale INAF 2023 (PI A. Vallenari) and from Prin INAF 2019 (PI
S. Lucatello).
\end{acknowledgements}

%\FloatBarrier

\begin{appendix}
\onecolumn
\section{References for the data on globular clusters}

In the following table we provide a synopsis of all the papers where the
original data on abundances can be retrieved for individual stars in GCs,
usually in electronic form at CDS.

\begin{table*}[h]
\centering
\caption{References for GC abundances used in this work}
\large
\begin{tabular}{l|c|c|c|c|c|c|}
\hline
GC    & [O/Fe] & [Mg/Fe] & [Si/Fe] & [Ca/Fe]  & [Ti/Fe]~{\sc i}& [Fe/H] \\
\hline
0104  & 1,2    & 2,3     &  2,3    &  4,5     &    5             &    6   \\
\hline
0288  & 1,2 & 2,5 & 2,5 & 4,5 & 5 & 6 \\
\hline
0362  & 7  &  7  &  7  &  7 &  7 & 7 \\
\hline
1851  & 8  &  8 &   8  &  8 &  8 & 8 \\
\hline
1904 & 1,2 & 2,5 & 2,5 & 4,5 & 5 & 6 \\
\hline
2808  & 9  &  9  &  9  &  9 &  9 & 9 \\
\hline
3201 & 1,2 & 2,5 & 2,5 & 4,5 & 5 & 6 \\
\hline
4590 & 1,2 & 2,5 & 2,5 & 4,5 & 5 & 6 \\
\hline
4833 & 10 &  10 &  10 &  10 &  10 &10 \\
\hline
5634 & 11 &  11 &  11 &  11 &  11 & 11 \\
\hline
5904 & 1,2 & 2,5 & 2,5 & 4,5 & 5 & 6 \\
\hline
6093 & 12 &  12 &  12 &  12 & 12 & 12 \\
\hline
6121 & 1,2 & 2,3 & 2,3 & 4,5 & 5 & 6 \\
\hline
6171 & 1,2 & 2,5 & 2,5 & 4,5 & 5 & 6 \\
\hline
6218 & 2,13 & 2,5 & 2,5 & 4,5 & 5 & 6 \\
\hline
6254 & 1,2 & 2,5 & 2,5 & 4,5 & 5 & 6 \\
\hline
6388 & 14 &  14 &  14 &  14 & 14 & 15 \\
\hline
6397 & 1,2 & 2,5 & 2,5 & 4,5 & 5 & 6 \\
\hline
6441 & 16 &  16 &  16 &  16 &  16 & 16 \\
\hline
6535 & 17 &  17 &  17 &  17 &  17 & 17 \\
\hline
6715 & 18 &  18 &  18 &  18 &  18 & 18 \\
\hline
6752 & 2,19& 2,20 & 2,20& 4,5 & 5 & 6 \\
\hline
6809 & 1,2 & 2,5 & 2,5 & 4,5 & 5 & 6 \\
\hline
6838 & 1,2 & 2,5 & 2,5 & 4,5 & 5 & 6 \\
\hline
7078 & 1,2 & 2,5 & 2,5 & 4,5 & 5 & 6 \\
\hline
7099 & 1,2 & 2,5 & 2,5 & 4,5 & 5 & 6 \\
\hline
Ter8  & 21 &  21 &  21 &  21 &  21& 21 \\
     
\hline
\end{tabular}

\label{t:tableA2ref}
\begin{list}{}{}
\item[-] 1=Carretta et al. (2009b), 2=Carretta et al. (2009a), 3=Carretta et al. (2013a)
\item[-] 4=Carretta et al. (2010b), 5=Carretta (2026, in press), 6=Carretta et al. (2009c)
\item[-] 7=Carretta et al. (2013b), 8=Carretta et al. (2011), 9=Carretta (2015)
\item[-] 10=Carretta et al. (2014a), 11=Carretta et al. (2017), 12=Carretta et al. (2015)
\item[-] 13=Carretta et al. (2007a), 14=Carretta and Bragaglia (2023), 15=Carretta and Bragaglia (2022)
\item[-] 16=Gratton et al. (2006,2007), 17=Bragaglia et al. (2017), 18=Carretta et al. (2010c)
\item[-] 19=Carretta et al. (2007b), 20=Carretta et al. (2012), 21=Carretta et al. (2014b)
\item[-] Based on observations collected at ESO telescopes under programmes
072.D-0507, 073.D-0211, 073.D-0.760, 081.D-286, 381.D-0329, 083.D-0208, 
085.D-0205, 087.B-0086, 093.B-0.583, 095.D-0834, 099.D-0047. 
\end{list}

\end{table*}

\end{appendix}

\end{document}